 \documentclass[smallabstract,smallcaptions]{dccpaper}

\usepackage{epsfig}
\usepackage{citesort}
\usepackage{amsmath}
\usepackage{amssymb}
\usepackage{color}
\usepackage{url}
\usepackage{subfigure}
\usepackage{threeparttable}
\usepackage{multirow}

\newlength{\figurewidth}
\newlength{\smallfigurewidth}

\begin{document}

\title
{\large
\textbf{A Diffusion Model Based Quality Enhancement Method for HEVC Compressed Video}
}

\author{%
Zheng Liu$^{\ast}$, Honggang Qi$^{\ast}$\\[0.5em]
{\small\begin{minipage}{\linewidth}\begin{center}
\begin{tabular}{ccc}
$^{\ast}$University of Chinese Academy of Sciences, Beijing, China. & \hspace*{0.5in} \\
$^{\ast}$\url{liuzheng18@mails.ucas.edu.cn}, $^{\ast}$\url{hgqi@ucas.ac.cn}\\
\end{tabular}
\end{center}\end{minipage}}
}


\maketitle
\thispagestyle{empty}

\begin{abstract}
Video post-processing methods can improve the quality of compressed videos at the decoder side. Most of the existing methods need to train corresponding models for compressed videos with different quantization parameters to improve the quality of compressed videos. However, in most cases, the quantization parameters of the decoded video are unknown. This makes existing methods have their limitations in improving video quality. To tackle this problem, this work proposes a diffusion model based post-processing method for compressed videos. The proposed method first estimates the feature vectors of the compressed video and then uses the estimated feature vectors as the prior information for the quality enhancement model to adaptively enhance the quality of compressed video with different quantization parameters. Experimental results show that the quality enhancement results of our proposed method on mixed datasets are superior to existing methods.
\end{abstract}

\Section{Introduction}
With the development of communication and multimedia technologies, he use of high-resolution, wide colour range video has brought about a clear and more realistic visual experience. However, these videos have led to an explosion in the amount of data in video files. According to Cisco's latest white paper \cite{CiscoVNI}, video data generates more than 80 \% of all traffic across the Internet by 2023. This explosive growth in data creates a challenge for video storage and transmission. To face this challenge, Joint Video
Exploration Team (JVET) has developed different video encoding standards \cite{h264avc,h265hevc,h266vvc}. By encoding video into binary bits files, the size of the video is reduced exponentially, easing the pressure on video storage and transmission.

Due to the introduction of some coding tools, compression artefacts are generated during video coding. These compression artefacts can seriously affect the quality of the video and have an impact on subsequent computer vision tasks. To mitigate the compression artifacts, in-loop filtering methods such as deblocking filtering (DBF) \cite{deblock2012}, sample adaptive offset filtering (SAO) \cite{sao2012}, and adaptive loop filtering (ALF) \cite{alf2013}are adopted in some new video coding standards. These in-loop filtering methods reduce the distortion of the reconstructed video by compensating the current region pixels with predefined rules. Although these in-loop filtering methods can effectively mitigate the compression artifact, they still have a few limitations.

In addition, many learning-based post-processing methods have been proposed to enhance video quality at the decoder side \cite{DaiVRCNN,WangDCAD2017,YangQECNN2019,YangMFQE12018,GuanMFQE22019,LuDeepKalman2019,DengSTDF2020}. These methods can be classified into two categories: video quality enhancement methods based on spatial domain information \cite{DaiVRCNN,WangDCAD2017} and video quality enhancement methods based on temporal domain information \cite{YangQECNN2019,YangMFQE12018,GuanMFQE22019,LuDeepKalman2019,DengSTDF2020}. The former reduces the pixel distortion based on the pixel information around the region to be enhanced. While the latter enhances a specific region by introducing temporal domain reference information to improve the video quality. Both types of methods do not require changes in the encoder structure and therefore do not introduce additional complexity.  Existing learning-based methods use compressed data with different quantization parameters to train the corresponding models, respectively. However, the quantization parameters of the video to be enhanced are often unknown at the decoder side. If the corresponding model is not used to process the video, it will cause the video quality enhancement results to fall short of expectations. To solve this problem, this work proposes a video quality enhancement method based on diffusion model. The method encodes the ground-truth image, the compressed image, and the quantization parameters into a feature vector and the quality of the compressed video is enhanced by this feature vector based decoder. The main contributions of this work are as follows:
\begin{itemize}
    \item In this work, we construct a quantization parameter-independent video quality enhancement model based on the estimator of the diffusion model. As we known, this is the first generative blind-QP video quality enhancement method based on the diffusion model.
    \item Our proposed method estimates the features of the input compressed video and uses the features as a prior information for quality enhancement. In this case, only one model needs to be trained to improve the quality of compressed videos with different quantization parameters.
\end{itemize}

The rest of the work is organised as follows. Section 2 briefly introduces several research works related to this work; Section 3 describes our proposed method in detail, including the implementation details of the different network models as well as the theoretical foundations; Section 4 demonstrates the video quality enhancement results; and Section 5 summarises the full work.

\Section{Related Works}
\subsection*{Video Quality Enhancement}
At present, three in-loop filtering methods are adopted in video coding standard \cite{deblock2012,sao2012,alf2013}. These in-loop filtering methods use the pixels of neighbouring coding units to mitigate compression artifacts. To further improve the coding efficiency, a large number of learning-based in-loop filtering methods have been proposed \cite{ZhangRHCNN2018,2019Content,9418683}. In \cite{ZhangRHCNN2018}, a model named RHCNN is applied after SAO module in HEVC and improve the coding performace in HEVC. In addition, some coding information during the video coding process can also be used as a priori information  \cite{2019Content,9418683}. In addition, there have been many research works proposing the use of post-processing methods to enhance the video quality at the decoder side \cite{WangDCAD2017,YangQECNN2019,YangMFQE12018,GuanMFQE22019,LuDeepKalman2019,DengSTDF2020}. \cite{WangDCAD2017,YangQECNN2019} Enhance the quality of compressed video by using correlation of spatial pixels in video.
Next, many researchers proposed to use multi-frame fusion quality enhancement methods based on temporal domain information \cite{YangMFQE12018,GuanMFQE22019,LuDeepKalman2019,DengSTDF2020}. In \cite{YangMFQE12018} and \cite{GuanMFQE22019}, features of different reference frames are extracted and then aligned for enhancing the quality of the current frame of video through a neural network model. \cite{DengSTDF2020} proposes a neural network model STDF based on deformable convolution that can automatically extract and align the feature information of different reference frames. This model achieves better quality enhancement results than previous models. All of the above methods require training the corresponding models for compressed videos with different quantization parameters. However, in practice, the quantization parameters of the decoded video are usually unknown, which limits the use of these methods in practical scenarios.
\subsection*{Image Generation with Diffusion Models}
Diffusion models \cite{ho2020denoising} have made advances in the field of image synthesis \cite{ho2020denoising,shi2022divae,10233841} as well as image restoration \cite{9954640,10214566,xia2023diffir}. Diffusion models can generate high-quality images with specified content. Through multiple iterations, the diffusion model achieved better results than other generative models such as GAN \cite{goodfellow2014generative}. \cite{9954640,10214566} propose image restoration or super-resolution methods based on diffusion models, but the overall efficiency of the models is not high due to the large number of iterations. In \cite{xia2023diffir}, an efficient diffusion model is proposed for image restoration. The research improved the performance of the model for image restoration by transforming the input image into a compact feature representation through the diffusion model. Inspired by this, we use a diffusion model to estimate the feature vectors of compressed videos with different quantization parameters. And the quality of compressed video with different quantization parameters is improved according to the feature vectors.
\Section{Proposed Method}
\subsection*{Diffusion Models}
Diffusion models \cite{ho2020denoising} is a method used to synthesise high quality images which is mainly divided into forward process and reverse process. In the forward process, a given image $x_0$ generates a fully Gaussian noise image $x_T$ by $T$ iterations. Each iteration can be represented by the following equation:
\begin{equation}\label{eq:forwardprocess}
    q(x_t|x_{t-1}) = \mathcal{N}(x_t; \sqrt{1-\beta_t}x_{t-1}, \beta_t\boldsymbol{I}),
\end{equation}
where $x_t$ is noise image at timestep $t$, $\mathcal{N}$ is the Gaussian distribution and $\beta_t$ is the pre-defined noise factor. $\boldsymbol{I}$ is the unit matrix. It is worth noting that the sample $x_t$ at arbitrary timestep $t$ in the forward process can be obtained by using initial image $x_0$, which can be expressed as follows:
\begin{equation}\label{eq:sample}
    q(x_t|x_0) = \mathcal{N}(x_t; \sqrt{\bar{\alpha_t}}x_0, (1-\bar{\alpha_t})\boldsymbol{I}),
\end{equation}
where $\alpha_t = 1-\beta_t$ and $\bar{\alpha}_t = \prod \limits_{i=1}^t \alpha_i$.

In the reverse process, the diffusion model first samples a Gaussian noise map $x_T$ and then gradually removes noise from $x_T$ until it produces a high quality output $x_0$, which can be expressed as:
\begin{equation}\label{eq:reverse}
    p(x_{t-1}|x_t, x_0) = \mathcal{N}(x_{t-1}; \mu_t(x_t, x_0), \sigma_{t}^2\boldsymbol{I}),
\end{equation}
where $\mu_t(x_t, x_0) = \frac{1}{\sqrt{\bar{\alpha_t}}}(x_t - \epsilon\frac{1-\alpha_t}{\sqrt{1-\bar{\alpha}_t}})$ and $\sigma_t^2 = \frac{1-\bar{\alpha}_{t-1}}{1-\bar{\alpha}_t}$. $\epsilon$ is the noise in $x_T$. In the reverse process, only the noise $\epsilon$ is uncertain and need to be estimated by a neural network. Assuming that the estimated network is $\epsilon_{\theta}(x_t,t)$, the diffusion model generates a noise map $x_t$ by randomly sampling timestep $t$ and noise $\epsilon$ according to Eq. (\ref{eq:sample}), and network is optimised by the gradient descent method according to the following manner \cite{ho2020denoising}:
\begin{equation}\label{eq:optimize}
    \nabla_\theta||\epsilon - \epsilon_{\theta}(\sqrt{\bar{\alpha}_t}x_0+\sqrt{1-\bar{\alpha}_t},t)||_2^2.
\end{equation}
\subsection*{Diffusion Based Video Quality Enhancement Model}
\begin{figure}[t]
\begin{center}
 \includegraphics[width = 1\textwidth]{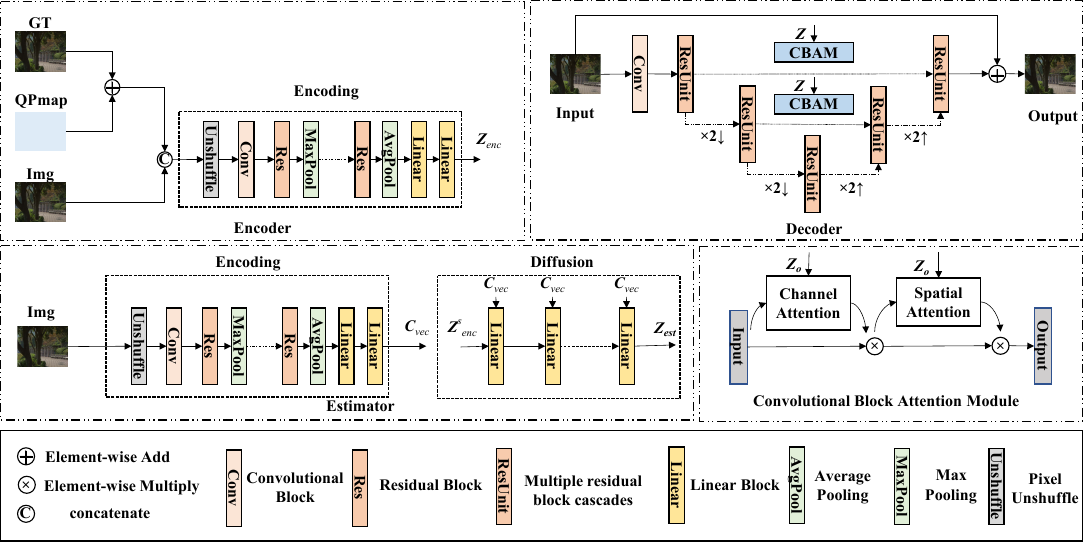}
\end{center}
\vspace{-0.2in}
\caption{\label{fig:1}%
Overview of the proposed model. The model can be divided into three modules: encoder, decoder and estimator. The encoder encodes ground-truth image, compressed image and QPmap as feature vector $Z_{est}$ and then the decoder generates the image based on $Z_{est}$ and compressed image. The estimator is used to estimate the feature vector $Z_{est}$. In the testing phase, we use only the estimator and decoder to generate high quality images.}
\end{figure}
In this work, we propose a diffusion based convolutional neural network model to improve the quality of HEVC compressed video. The overview of the model is shown in Figure \ref{fig:1}. Inspired by \cite{xia2023diffir}, our model consists of three parts: encoder, decoder and estimator.

In the encoder, the inputs are the ground-truth image (GT), the quantization parameter map (QPmap) and the compressed image (Img). We first add the QPmap to the ground-truth image and then concatenate it together with the compressed image to form the input map $x_{in}$, which can be expressed as follows:
\begin{equation}
    x_{in} = cat(Img,(GT \oplus QPmap)).
\end{equation}
Then the pixel input map $x_{in}$ is shuffled and encoded by a convolutional neural network. During the encoding process, we increase the number of channels in the feature map by adjusting the number of channels in the convolutional kernel of the convolutional neural network, and reduce the spatial size of the feature map by max-pooling. After a global average pooling layer, the feature map is used as input to the linear layer and finally encoded as a fixed-length feature vector $\textbf{Z}_{enc}$. The output $\textbf{Z}_{enc}$ of the encoder contains information about the ground-truth image as well as information about the quantization parameters of the encoding.

In the decoder, we use Unet as a backbone network to enhance the quality of compressed video. Unet is divided into two processes: down-sampling and up-sampling. In the process of down-sampling, the input feature map is down-sampled by max-pooling and the feature information of different scales is continuously extracted through the residual units. After the down-sampling process, the model will gradually fuse the feature information of the current and previous levels. In particular, in the process of feature fusion, we employ a convolutional block attention module (CBAM) \cite{woo2018cbam} embedded with  feature vector $\textbf{Z}_{enc}$ to enhance the forward features. Given a feature map $\textbf{F} \in R^{C\times H \times W}$, CBAM processes input feature maps through a channel attention operation $\textbf{M}_c$ and a spatial attention operation $\textbf{M}_s$. This process can be summaried as:
\begin{equation}\label{eq:cbam}
\textbf{F}_c = \textbf{M}_c(\textbf{F})\otimes \textbf{F},
\textbf{F}_{sc} = \textbf{M}_s(\textbf{F}_c)\otimes \textbf{F}_c,
\end{equation}
where $\otimes$ represents element-wise multiply. After that, the channel attention operation $\textbf{M}_c$ can be calculated by the following equation:
\begin{equation}\label{eq:mc}
\textbf{M}_c(\textbf{F}) = \sigma (MLP(AvgPool(\textbf{F}))+MLP(MaxPool(\textbf{F}))+MLP(\textbf{Z})),  
\end{equation}
where $\sigma$ denotes normalisation function, $MLP$ denotes the mapping function of linear layers and $\textbf{Z}$ is feature vector generated by encoder or estimator. In addition, the spatial attention operation can be obtained by the following equation:
\begin{equation}\label{eq:ms}
\textbf{M}_s(\textbf{F}) = \sigma (f_{conv}(cat(AvgPool(\textbf{F}), MaxPool(\textbf{F}), \textbf{Z}))),  
\end{equation}
where $\sigma$ denotes normalisation function, $f_{conv}(\cdot)$ denotes mapping function of convolutional layer and $cat$ denotes concatenate at channel-wise. 

Finally, fused features from different scales are processed through convolutional layers to generate a reconstructed residual image $x_{res}$ and added to the input image $x_{in}$ to generate a high-quality reconstructed image $x_{out}$.
\begin{equation}
    x_{out} = x_{in} + x_{res}.
\end{equation}

In the estimator, we first encode compressed image into a conditional vector $\textbf{C}_{vec}$. Then we generate Gaussian noise feature vectors $\textbf{Z}_{enc}^n$ based on the sampling feature vector $\textbf{Z}_{enc}^s$, condition vector $\textbf{C}_{vec}$, and timestep $t$. According to Eq. (\ref{eq:optimize}), the noise at each time step can be estimated through the network.
Afterwards, we estimate the noise $\epsilon$ through the network and remove it through $T$ iterations, ultimately obtaining an accurate estimate of the feature vector $\textbf{Z}_{est}$. $\textbf{Z}_{est}$ can guide the decoder to generate high-quality reconstructed images. The training methods for encoders, decoders, and estimators will be described in the next section.

\Section{Experiment Results}
\subsection*{Training Settings}
We use BSDS500 \cite{BSDS500} as the training set, while the 17 test sequences recommend by HEVC form the testing set. All the images are compressed by HM reference software with AI(All Intra) mode at QP$\in\{27,32,37,42\}$. Finally the video data compressed with different quantization parameters will be blended together to form the dataset. Each sample in the dataset contains the ground-truth image, compressed image, and corresponding quantization parameter values.

In the training stage, we first jointly train the encoder and decoder. The luminance channel in YCbCr space of the compressed images is extractedfor training the model. The loss function of the training process is defined as follows: 
\begin{equation}\label{eq:mse}
 L_{rec} =||x_{GT} - x_{out}||_2^2.  
\end{equation}
Then, we train the estimator based on the output of the encoder. The loss function of the estimator is defined as follows:
\begin{equation}\label{eq:zz}
 L_{est} =||\textbf{Z}_{est} - \textbf{Z}_{enc}||_2^2.  
\end{equation}

In the testing stage, we first randomly sample Gaussian noise $\textbf{Z}$. Then, the estimator generates a feature vector $\textbf{Z}_{est}$ through $T$ iterations based on the condition vector $C_{vec}$ generated from the compressed image. Finally, the decoder generates high-quality image based on compressed image and feature vector $\textbf{Z}_{est}$.
 \subsection*{Discussion}
In this section, we retrain the model from \cite{DaiVRCNN,WangDCAD2017,YangQECNN2019} on our train set and compare the performance of the models. $\Delta PSNR$ is used to evaluate the performance of the model. $\Delta PSNR$ can be calculated as follows:
\begin{equation}\label{eq:psnr}
    \Delta PSNR = PSNR_{EN} - PSNR_{HM},
\end{equation}
where $PSNR_{EN}$ and $PSNR_{HM}$ denote the PSNR values for enhanced and compressed video, respectively. The results are list in Table \ref{tab:tab1}.
\begin{table}[!t]
\centering
\caption{Quality enhancement results of different models}
\label{tab:tab1}
\begin{tabular}{cccccc}
\hline
\multirow{2}{*}{Class} & \multirow{2}{*}{SequenceName} & \multicolumn{4}{c}{$\Delta$PSNR(dB$\uparrow$)} \\ \cline{3-6} 
 &  & Dai \cite{DaiVRCNN} & Wang \cite{WangDCAD2017} & Yang \cite{YangQECNN2019} & Ours \\ \hline
\multirow{2}{*}{\begin{tabular}[c]{@{}c@{}}ClassA\\ 2560$\times$1440\end{tabular}} & Traffic & 0.075 & 0.066 & 0.094 & \textbf{0.343} \\ \cline{2-6} 
 & PeopleOnStreet & -0.049 & -0.157 & -0.018 & \textbf{0.258} \\ \hline
\multirow{5}{*}{\begin{tabular}[c]{@{}c@{}}ClassB\\ 1920$\times$1080\end{tabular}} & BasketballDrive & 0.024 & 0.018 & 0.049 & \textbf{0.142} \\ \cline{2-6} 
 & BQTerrace & 0.055 & 0.067 & 0.051 & \textbf{0.241} \\ \cline{2-6} 
 & Cactus & 0.034 & -0.126 & 0.080 & \textbf{0.178} \\ \cline{2-6} 
 & Kimono & 0.100 & 0.105 & 0.117 & \textbf{0.228} \\ \cline{2-6} 
 & ParkScene & 0.104 & 0.137 & 0.115 & \textbf{0.216} \\ \hline
\multirow{3}{*}{\begin{tabular}[c]{@{}c@{}}ClassC\\ 832$\times$480\end{tabular}} & BasketballDrill & 0.155 & 0.217 & 0.210 & \textbf{0.305} \\ \cline{2-6} 
 & BQMall & 0.117 & 0.160 & 0.116 & \textbf{0.277} \\ \cline{2-6} 
 & PartyScene & 0.104 & 0.160 & 0.083 & \textbf{0.192} \\ \hline
\multirow{4}{*}{\begin{tabular}[c]{@{}c@{}}ClassD\\ 416$\times$240\end{tabular}} & BasketballPass & 0.095 & 0.214 & 0.123 & \textbf{0.243} \\ \cline{2-6} 
 & RaceHorses & 0.222 & 0.273 & 0.224 & \textbf{0.389} \\ \cline{2-6} 
 & BlowingBubbles & 0.101 & 0.163 & 0.108 & \textbf{0.181} \\ \cline{2-6} 
 & BQSquare & 0.182 & 0.301 & 0.108 & \textbf{0.358} \\ \hline
\multirow{3}{*}{\begin{tabular}[c]{@{}c@{}}ClassE\\ 1280$\times$720\end{tabular}} & Johnny & 0.188 & 0.200 & 0.199 & \textbf{0.231} \\ \cline{2-6} 
 & FourPeople & 0.292 & 0.252 & 0.294 & \textbf{0.386} \\ \cline{2-6} 
 & KristenAndSara & 0.248 & 0.111 & 0.243 & \textbf{0.349} \\ \hline
\multicolumn{2}{c}{Average} & 0.120 & 0.127 & 0.129 & \textbf{0.266} \\ \hline
  \multicolumn{6}{c}{$\Delta$PSNR denotes the difference between output image and HM image.}\\
  \multicolumn{6}{c}{Higher values indicate better performance. The black font is the best.}\\
\end{tabular}
\end{table}

As shown in Table \ref{tab:tab1}, our model improves quality by 0.266 dB on average on the test set, which is higher than existing models. And the video quality can be improved by 0.142dB $\sim$ 0.389dB on average on different test sequences. The most significant quality enhancement result is found on the RaceHorses sequence, which reaches 0.389 dB. These indicate that our model adaptively improves the quality of compressed videos with different quantization parameters. And as can be seen from columns 3 to 5 of Table \ref{tab:tab1}, the existing models have limited quality enhancement results on the mixed dataset. In particular, on the PeopleOnStreet sequence, the existing model decreases the video quality on the mixed dataset. This may be due to the fact that in video coding, videos compressed with different quantization parameters have different features, and in the absence of sufficient a prior information, the existing models cannot learn these compression features well, and hence have limited quality enhancement performance. And our proposed model first estimates the features $\textbf{Z}_{est}$ of the current video. The estimated feature vector $\textbf{Z}_{est}$ contains a large amount of input video information and can be used as a prior information for the current video to enhance the video quality. With this method, the model fully learns the feature information of compressed videos with different quantization parameters to adaptively enhance the quality of compressed videos with different quantization parameters, and achieves good quality enhancement results on the mixed dataset. 
\subsection*{Ablation Study}
In order to verify the reasonableness of our model design, in this section we perform ablation experiments and present the results in Table \ref{tab:tab2}.
\begin{table}[!t]
\centering
\caption{Ablation results}
\label{tab:tab2}
\begin{tabular}{cclll}
\hline
\multirow{2}{*}{Class} & \multirow{2}{*}{SequenceName} & \multicolumn{3}{c}{$\Delta$PSNR(dB$\uparrow$)} \\ \cline{3-5} 
 &  & \multicolumn{1}{c}{NoEst} & \multicolumn{1}{c}{NoDiff} & \multicolumn{1}{c}{Ours} \\ \hline
\multirow{2}{*}{\begin{tabular}[c]{@{}c@{}}ClassA\\ 2560$\times$1440\end{tabular}} & Traffic & 0.001 & 0.295 & \textbf{0.343} \\ \cline{2-5} 
 & PeopleOnStreet & -0.357 & 0.207 & \textbf{0.258} \\ \hline
\multirow{5}{*}{\begin{tabular}[c]{@{}c@{}}ClassB\\ 1920$\times$1080\end{tabular}} & BasketballDrive & -0.187 & 0.033 & \textbf{0.142} \\ \cline{2-5} 
 & BQTerrace & 0.005 & 0.189 & \textbf{0.241} \\ \cline{2-5} 
 & Cactus & -0.067 & 0.081 & \textbf{0.178} \\ \cline{2-5} 
 & Kimono & 0.183 & 0.206 & \textbf{0.228} \\ \cline{2-5} 
 & ParkScene & 0.172 & 0.181 & \textbf{0.216} \\ \hline
\multirow{3}{*}{\begin{tabular}[c]{@{}c@{}}ClassC\\ 832$\times$480\end{tabular}} & BasketballDrill & 0.301 & 0.204 & \textbf{0.305} \\ \cline{2-5} 
 & BQMall & 0.151 & 0.195 & \textbf{0.277} \\ \cline{2-5} 
 & PartyScene & 0.150 & 0.186 & \textbf{0.192} \\ \hline
\multirow{4}{*}{\begin{tabular}[c]{@{}c@{}}ClassD\\ 416$\times$240\end{tabular}} & BasketballPass & 0.188 & 0.212 & \textbf{0.243} \\ \cline{2-5} 
 & RaceHorses & 0.336 & 0.289 & \textbf{0.389} \\ \cline{2-5} 
 & BlowingBubbles & 0.171 & 0.165 & \textbf{0.181} \\ \cline{2-5} 
 & BQSquare & 0.332 & \textbf{0.372} & 0.358 \\ \hline
\multirow{3}{*}{\begin{tabular}[c]{@{}c@{}}ClassE\\ 1280$\times$720\end{tabular}} & Johnny & 0.150 & 0.202 & \textbf{0.231} \\ \cline{2-5} 
 & FourPeople & -0.177 & 0.362 & \textbf{0.386} \\ \cline{2-5} 
 & KristenAndSara & -0.458 & 0.253 & \textbf{0.349} \\ \hline
\multicolumn{2}{c}{Average} & 0.053 & 0.214 & \multicolumn{1}{c}{\textbf{0.266}} \\ \hline
\multicolumn{5}{c}{$\Delta$PSNR denotes the difference between output image} \\
\multicolumn{5}{c}{ and HM image. The black font is the best.}
\end{tabular}
\end{table}
Column 3 of Table \ref{tab:tab2} shows the quality enhancement results of the decoder without the feature vector $\textbf{Z}_{est}$. At this time the model can only improve the average quality on the mixed dataset by 0.053 dB. It also degrades the video quality on several test sequences. This shows that the decoder has more limited quality enhancement results on the mixed dataset without the video feature vector $\textbf{Z}_{est}$. Column 4 of Table 2 shows the video quality enhancement results of the decoder using the video feature vectors $\textbf{Z}_{est}$ directly estimated by the neural network model as input. In this case, the model can improve the average quality on the mixed dataset by 0.214 dB. This result is still lower than our proposed diffusion model based video quality enhancement method. It indicates that using the diffusion model can estimate the feature vectors $\textbf{Z}$ of the video more accurately and achieve better quality enhancement results on the mixed dataset.
\Section{Conclusion}
In this work, we build a hybrid dataset containing compressed videos with different quantization parameters. Moreover, we propose a video post-processing method based on the diffusion model to improve the quality of compressed videos. The model first accurately estimates the feature information of the input video based on the diffusion model. Then it improves the quality of the input video based on the feature information. Experimental results show that our proposed method can improve the quality of compressed video with different quantization parameters. The performance on mixed datasets is better than the existing methods.
\Section{References}
\bibliographystyle{IEEEbib}
\bibliography{refs}

\end{document}